\documentclass[12pt]{article}
\usepackage{listings}
\usepackage{color}
\usepackage[utf8]{inputenc}
\usepackage{pgf,tikz}
\usetikzlibrary{arrows}
\definecolor{dkgreen}{rgb}{0,0.6,0}
\definecolor{gray}{rgb}{0.5,0.5,0.5}
\definecolor{mauve}{rgb}{0.58,0,0.82}

\begin{document}
\textbf{A Simple Circle Discretization Algorithm With Applications}\\
			\begin{center}
			Craig Vercueil\\
			\end{center}

Abstract.In CNC manufacturing,there often arises the need to create G-Code programs which require the calculation of discrete x-y coordinate point pairs(2D).
An example of this situation is when the programmer needs to create a program to machine a helix(or thread).The required toolpath will be a set of points on a helix curve.The problem now entails calculating the number of points along this curve.If there are too few points,the toolpath will not be smooth.Too many points and the program becomes too big.This article will serve to provide a simple way to divide a circle into discrete points,with a notion of "dimensional tolerance" built into the algorithm.\\

\textbf{Introduction}

When required to machine a helix or thread using CNC machines,it is sometimes necessary to calculate discrete point pairs on this curve.The cutting tool must follow a path along these points.
We view the helix as a circle in the X-Y Cartesian plane with an advance in the Z-axis direction.The problem now facing the CNC-programmer is to determine the number of points for a smooth toolpath(sometimes a rough toolpath may also be needed).What follows is a simple non-arbitrary method to calculate the number of points.\\

\textbf{The Method}

We will use a well known limiting argument to briefly discuss the ideas behind the technique.
First we approximate the number of points on the circle by thinking of an inscribed regular polygon P with \textit{n} vertices.As \textit{n} $\rightarrow \infty$, the inscribed polygon approaches the bounding circle C.Similarly,as \textit{n} $\rightarrow \infty$,the area A$_{p}$ of the inscribed polygon P approaches the area of our bounding circle A$_{c}$.That is,as \textit{n} $\rightarrow \infty$, A$_{p}\rightarrow$ A$_{c}$.

The area A$_{c}$ = $\pi$r$^{2}$ and the area A$_{p}$ = 2\textit{n}(\textit{ah}).In the latter formula for A$_{p}$,\textit{a} is the apothem of P,and \textit{h} is the height of a right angled triangle formed by \textit{r},\textit{a} and \textit{h}.See Fig1.It is obvious that if we need a finite value for\textit{ n},then \textit{a} can never equal \textit{r}, i.e. \textit{a}$\neq$\textit{r} and \textit{r} is an upper limit for \textit{a}.Now we can encode a notion of "dimensional tolerance" in our calculation, by choosing how close we want \textit{a} to get to \textit{r}.Let us denote this difference of \textit{r-a},by $\delta$.By setting 
$\delta$ to some small posive value(eg. $\delta$ = 0.1),we generate an inscribed polygon with apothem \textit{a} = \textit{(r - 0.1)} or genrally \textit{a} = \textit{r} - $\delta$.From this we can now calculate
the value of \textit{n}.Since \textit{r} is given we have:
 the height \textit{h} = $\sqrt{r^{2}-a^{2}}$.
 The angle between \textit{r} and \textit{a} is $\theta$ = Arctan\textit{(h/a)} in radians.
 Putting this all together we have, \textit{n} = $\pi$/$\theta$.In a step-by-step
 procedure:\\
 1).Choose a value for $\delta$ and set tolerance.\\
 2).Use theorem of Pythagoras to calculate \textit{h} = $\sqrt{r^{2}-a^{2}}$.\\
 3).Use $\theta$ = Arctan\textit{(h/a)} to calculate $\theta$.\\
 4).Calculate value of \textit{n} = $\pi$/$\theta$.\\

 \textbf{     }
 
\textbf{The Application}

Once we have a value for \textit{n} we can use this value in our computations.This is best illustrated by an example.We will use Java as our language for this purpose:Listing 1.
\begin{lstlisting}
import java.util.*;

public class NumCalc {
	/**
	 * @param args
	 */
	public static void main(String[] args) {

		// declare variables
		double radius;
		double angle;
		double theta;
		double chord;
		double height;
		double apothem;
		double numpts;
		double PI = 3.14159265;

		Scanner in = new Scanner(System.in);

		System.out.println("Enter a value for the radius");
		radius = in.nextDouble();

		double dia = 2 * radius;
		apothem = radius - 0.0005;// chosen value for tolerance
		height = Math.sqrt((radius * radius) - (apothem * apothem));
		chord = 2 * height;
		theta = Math.atan(height / apothem);
		angle = 2 * (theta);// angle between points
		numpts = (2 * PI) / angle;// calculate number of points

		int n = ((int) numpts) + 1;//convert to int
		System.out.println("The number of points is " + n);
		System.out.printf("The angle is %01.3f\r\n", angle);
	}
}
\end{lstlisting}
Listing 1 above shows the code to generate a value for \textit{n}.This code can be implemented in another Java class which is our actual goal:

Listing 2. This is a full application employing our algorithm.This code can be copied and run on Eclipse.The resulting G-Codes can be used on a CNC 3-Axis Vertical Milling Machine.The G-Codes will be created in a file called "TestHelix.nc".
\begin{lstlisting}
import java.util.*;
import java.io.*;

public class Spiral {

	/**
	 * @param args
	 */
	public static void main(String[] args) {
		{
			try {

				FileWriter outFile = new FileWriter("TestHelix.nc"); // create
																	// file
				PrintWriter out = new PrintWriter(outFile);

				Scanner in = new Scanner(System.in);

				// variables
				double revs;
				int plungefeed;
				double theta, startang, angle, angl;
				double PI = 3.14159265;
				double cutrad, cutdia;
				double PCD, radius;
				double cbdia;
				double xcent, ycent;
				double numpts;
				double tradius;
				double tangle;
				double ttheta;
				double pitch;
				double apothem;
				double height;
				double length;
				double tolerance;

	// Calculate the number of holes and the angle theta
	// First step get inputs to calculate the number of points
	// get cutterdia
	System.out.println("Enter the cutter dia\n");
	cutdia = in.nextDouble();
	cutrad = cutdia / 2;

	// get center for X
	System.out.println("Enter the center for X\n");
	xcent = in.nextDouble();

	// get center for Y
	System.out.println("Enter the center for Y\n");
	ycent = in.nextDouble();

	// get the radius of the bore
	System.out.println("Enter a value for the radius");
	tradius = in.nextDouble();
	radius = tradius - cutrad;

	// get the pitch
	System.out.println("Enter the pitch\n");
	pitch = in.nextDouble();

	// get the length of the bore
	System.out.println("Enter the length of the bore\n");
	length = in.nextDouble();
	
	System.out.println("Enter the tolerance,a small value\n");
	tolerance = in.nextDouble();
	revs = length / pitch;
	int revsk = (int) revs;

 

	
	// Now calculate the number of points
	//this is our algorithm			
	apothem = radius - tolerance;
	height = Math.sqrt((radius * radius) - (apothem * apothem));
	theta = Math.atan(height / apothem);
	angle = 2 * (theta);// angle between points
	numpts = (2 * PI) / angle;// calculate number of points
	int n = ((int) numpts) + 1;// convert numpts to integer n

	System.out.println("The number of points is " + n);
	System.out.printf("The angle is %01.6f\r\n", angle);

	// Print program header
	out.println("%O1000");
	out.println("G54 G17 G40 G80 G90");
	out.println("S1000 M03");
	out.println("G01 Z100. F1000");
	out.println("X" + xcent + " Y" + ycent + " Z20.00 F3000");

	for (int k = 0; k < revsk; k++) // loop over revs

	{

	for (int i = 0; i < n; i++) // loop over number of points n

	{

 theta = (i * angle);

 double xval[] = new double[n]; // array for x coordinate
 double yval[] = new double[n]; // array for y coordinate
 double zval[] = new double[n]; // array for z coordinate

xval[i] = (radius * (Math.cos(theta))) + xcent; //x value
																		
yval[i] = (radius * (Math.sin(theta))) + ycent; //y value
																		
zval[i] = (k * pitch) + ((theta / (2 * PI)) * pitch); //z value
																			

System.out.printf("X%01.3f Y%01.3f Z%01.3f\r\n",
			xval[i], yval[i], zval[i]);

 out.printf("X%01.3f Y%01.3f Z%01.3f\r\n", xval[i],
			yval[i], zval[i]); // print the coordinates
					}
				}

out.println("X" + xcent + " Y" + ycent + " F1000");
out.println("Z100.000");
out.println("M30");
out.close(); // close output file

}
 catch (IOException e) // error handling
{
	e.printStackTrace();
		}
	}
}
}
\end{lstlisting}
\definecolor{uququq}{rgb}{0.25,0.25,0.25}
\definecolor{zzttqq}{rgb}{0.6,0.2,0}
\definecolor{qqqqff}{rgb}{0,0,1}
\begin{tikzpicture}[line cap=round,line join=round,>=triangle 45,x=1.0cm,y=1.0cm]
\clip(-9.98,-2.6) rectangle (12.91,14.47);
\fill[color=zzttqq,fill=zzttqq,fill opacity=0.1] (2,10) -- (-2,10) -- (-4,6.54) -- (-2,3.07) -- (2,3.07) -- (4,6.54) -- cycle;
\draw [color=zzttqq] (2,10)-- (-2,10);
\draw [color=zzttqq] (-2,10)-- (-4,6.54);
\draw [color=zzttqq] (-4,6.54)-- (-2,3.07);
\draw [color=zzttqq] (-2,3.07)-- (2,3.07);
\draw [color=zzttqq] (2,3.07)-- (4,6.54);
\draw [color=zzttqq] (4,6.54)-- (2,10);
\draw(0,6.54) circle (4cm);
\draw [domain=-9.98:12.91] plot(\x,{(--52.29-0*\x)/8});
\draw (0,-2.6) -- (0,14.47);
\draw (0,6.54)-- (2,10);
\draw (0,6.54)-- (3,8.27);
\draw (2,10)-- (3,8.27);
\draw (-4.38,4.19) node[anchor=north west] {Circle C};
\draw (1.41,5.32) node[anchor=north west] {Polygon P};
\draw (4.26,7) node[anchor=north west] {n=1};
\draw (2.33,10.33) node[anchor=north west] {n=2};
\draw (-2.39,10.64) node[anchor=north west] {n=3};
\draw (-4.79,7.08) node[anchor=north west] {n=4};
\draw (-2.5,2.88) node[anchor=north west] {n=5};
\draw (2.23,3.23) node[anchor=north west] {n=6};
\draw (0.37,7.43) node[anchor=north west] {$\theta$};
\draw (-6.61,12.77) node[anchor=north west] {Fig 1.};
\begin{scriptsize}
\fill [color=qqqqff] (2,10) circle (1.5pt);
\fill [color=qqqqff] (-2,10) circle (1.5pt);
\fill [color=uququq] (-4,6.54) circle (1.5pt);
\fill [color=uququq] (-2,3.07) circle (1.5pt);
\fill [color=uququq] (2,3.07) circle (1.5pt);
\fill [color=uququq] (4,6.54) circle (1.5pt);
\draw[color=black] (0.74,8.72) node {$radius$};
\fill [color=uququq] (3,8.27) circle (1.5pt);
\fill [color=uququq] (0,6.54) circle (1.5pt);
\draw[color=black] (2.13,7.43) node {$apothem$};
\draw[color=black] (2.68,9.1) node {$h$};
\end{scriptsize}
\end{tikzpicture}

\definecolor{uququq}{rgb}{0.25,0.25,0.25}
\definecolor{zzttqq}{rgb}{0.6,0.2,0}
\definecolor{qqqqff}{rgb}{0,0,1}
\begin{tikzpicture}[line cap=round,line join=round,>=triangle 45,x=1.0cm,y=1.0cm]
\clip(-8.54,-11.23) rectangle (25.12,13.55);
\fill[color=zzttqq,fill=zzttqq,fill opacity=0.1] (1,10) -- (-1,10) -- (-2.9,9.38) -- (-4.52,8.21) -- (-5.7,6.59) -- (-6.31,4.69) -- (-6.31,2.69) -- (-5.7,0.78) -- (-4.52,-0.83) -- (-2.9,-2.01) -- (-1,-2.63) -- (1,-2.63) -- (2.9,-2.01) -- (4.52,-0.83) -- (5.7,0.78) -- (6.31,2.69) -- (6.31,4.69) -- (5.7,6.59) -- (4.52,8.21) -- (2.9,9.38) -- cycle;
\draw [color=zzttqq] (1,10)-- (-1,10);
\draw [color=zzttqq] (-1,10)-- (-2.9,9.38);
\draw [color=zzttqq] (-2.9,9.38)-- (-4.52,8.21);
\draw [color=zzttqq] (-4.52,8.21)-- (-5.7,6.59);
\draw [color=zzttqq] (-5.7,6.59)-- (-6.31,4.69);
\draw [color=zzttqq] (-6.31,4.69)-- (-6.31,2.69);
\draw [color=zzttqq] (-6.31,2.69)-- (-5.7,0.78);
\draw [color=zzttqq] (-5.7,0.78)-- (-4.52,-0.83);
\draw [color=zzttqq] (-4.52,-0.83)-- (-2.9,-2.01);
\draw [color=zzttqq] (-2.9,-2.01)-- (-1,-2.63);
\draw [color=zzttqq] (-1,-2.63)-- (1,-2.63);
\draw [color=zzttqq] (1,-2.63)-- (2.9,-2.01);
\draw [color=zzttqq] (2.9,-2.01)-- (4.52,-0.83);
\draw [color=zzttqq] (4.52,-0.83)-- (5.7,0.78);
\draw [color=zzttqq] (5.7,0.78)-- (6.31,2.69);
\draw [color=zzttqq] (6.31,2.69)-- (6.31,4.69);
\draw [color=zzttqq] (6.31,4.69)-- (5.7,6.59);
\draw [color=zzttqq] (5.7,6.59)-- (4.52,8.21);
\draw [color=zzttqq] (4.52,8.21)-- (2.9,9.38);
\draw [color=zzttqq] (2.9,9.38)-- (1,10);
\draw(0,3.69) circle (6.39cm);
\draw [domain=-8.54:25.12] plot(\x,{(--46.55-0*\x)/12.63});
\draw (0,-11.23) -- (0,13.55);
\draw (-7.35,12.05) node[anchor=north west] {Fig 2. With n=20,};
\draw (2.51,2.41) node[anchor=north west] {Polygon P};
\draw (-7.76,0.13) node[anchor=north west] {Circle C};
\draw(3.68,7.2) circle (1.31cm);
\draw(2.01,8.37) circle (1.31cm);
\draw(0.88,8.64) circle (1.31cm);
\draw (0.22,6.14) node[anchor=north west] {Cutter Path};
\draw [shift={(0.5,4.46)}] plot[domain=0.71:1.48,variable=\t]({1*4.2*cos(\t r)+0*4.2*sin(\t r)},{0*4.2*cos(\t r)+1*4.2*sin(\t r)});
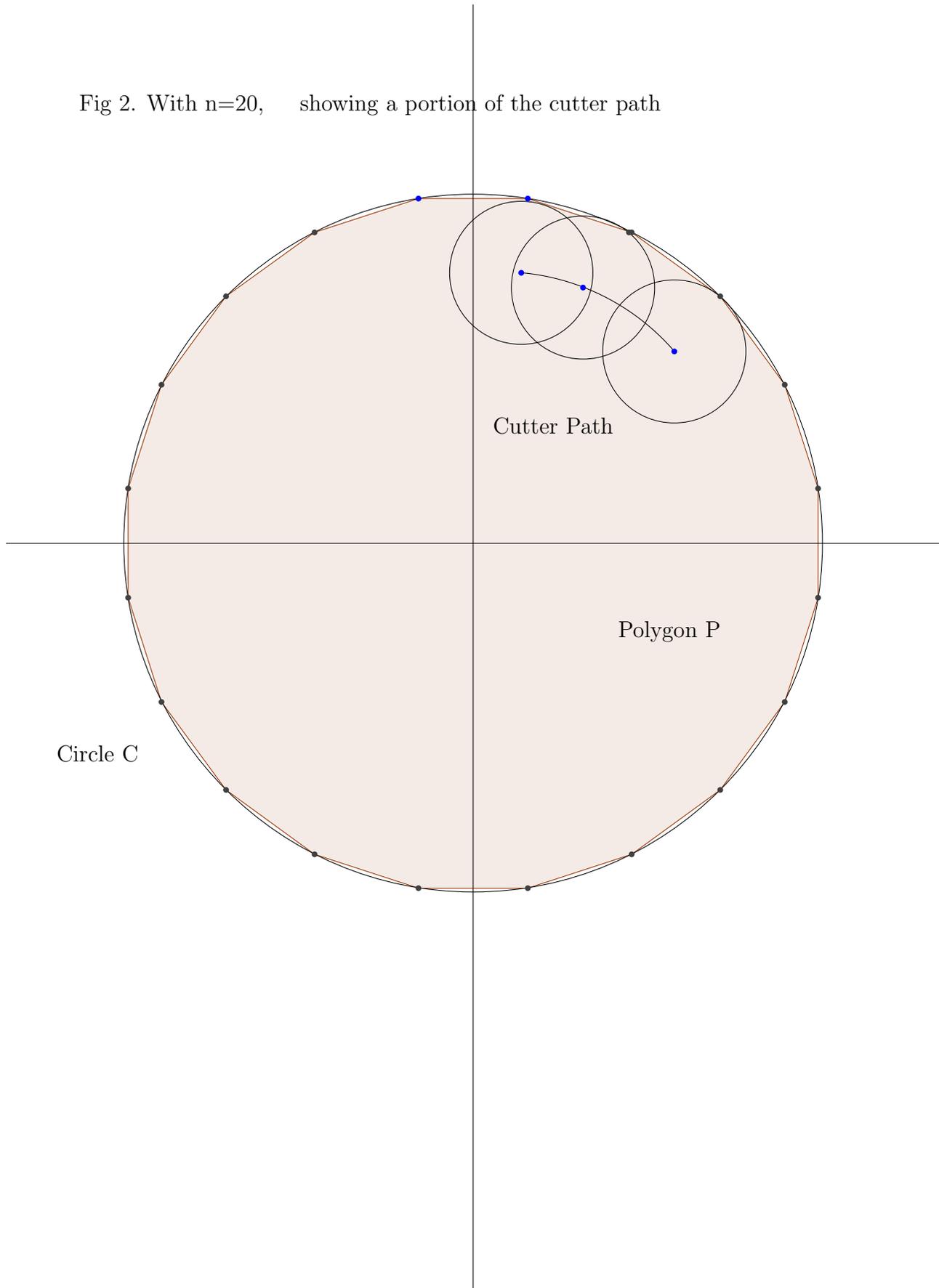
\draw (-3.31,12.05) node[anchor=north west] {showing a portion of the cutter path};
\begin{scriptsize}
\fill [color=qqqqff] (1,10) circle (1.5pt);
\fill [color=qqqqff] (-1,10) circle (1.5pt);
\fill [color=uququq] (-2.9,9.38) circle (1.5pt);
\fill [color=uququq] (-4.52,8.21) circle (1.5pt);
\fill [color=uququq] (-5.7,6.59) circle (1.5pt);
\fill [color=uququq] (-6.31,4.69) circle (1.5pt);
\fill [color=uququq] (-6.31,2.69) circle (1.5pt);
\fill [color=uququq] (-4.52,-0.83) circle (1.5pt);
\fill [color=uququq] (-2.9,-2.01) circle (1.5pt);
\fill [color=uququq] (-1,-2.63) circle (1.5pt);
\fill [color=uququq] (1,-2.63) circle (1.5pt);
\fill [color=uququq] (2.9,-2.01) circle (1.5pt);
\fill [color=uququq] (4.52,-0.83) circle (1.5pt);
\fill [color=uququq] (5.7,0.78) circle (1.5pt);
\fill [color=uququq] (6.31,2.69) circle (1.5pt);
\fill [color=uququq] (6.31,4.69) circle (1.5pt);
\fill [color=uququq] (5.7,6.59) circle (1.5pt);
\fill [color=uququq] (4.52,8.21) circle (1.5pt);
\fill [color=uququq] (2.9,9.38) circle (1.5pt);
\fill [color=uququq] (-5.7,0.78) circle (1.5pt);
\fill [color=qqqqff] (3.68,7.2) circle (1.5pt);
\fill [color=uququq] (2.85,9.38) circle (1.5pt);
\fill [color=qqqqff] (2.01,8.37) circle (1.5pt);
\fill [color=qqqqff] (0.88,8.64) circle (1.5pt);
\end{scriptsize}
\end{tikzpicture}
\textbf{Conclusion}\\
The algorithm as discussed above is useful in that it supplies the user
with a non-arbtrary method to discretize a circular toolpath curve.Once this value is computed it is used to generate G-Codes.The example Java code can be modified to suit spirals,helices,circles and even eliptical helices.Although there is CAD/CAM software available for creating such toolpaths,this is an inexpensive alternative.
\end{document}